# High thermal conductivity of hexagonal boron nitride laminates

Jin-Cheng Zheng[1,2*], Liang Zhang[1], A V Kretinin[3,4], S V Morozov[4,5,6], Yi Bo Wang[4], Tun Wang[7], Xiaojun Li[1], Fei Ren[1], Jingyu Zhang[8], Ching-Yu Lu[8], Jia-Cing Chen[8], Miao Lu[7], Hui-Qiong Wang[1,9], A K Geim[4] and K S Novoselov[4*]

[1]Department of Physics, Xiamen University, Xiamen 361005, China
[2]Fujian Provincial Key Laboratory of Theoretical and Computational Chemistry, Xiamen University, Xiamen 361005, China
[3]School of Materials, University of Manchester, Oxford Road, Manchester, M13 9PL, UK
[4]School of Physics and Astronomy, University of Manchester, Oxford Road, Manchester M13 9PL, UK
[5]Institute of Microelectronics Technology and High Purity Materials, RAS, Chernogolovka 142432, Russia
[6]National University of Science and Technology 'MISiS', 119049 Leninsky pr.4, Moscow, Russia
[7]MEMS research center, Pen-Tung Sah Institute of Micro-Nano Science & Technology, Xiamen University, Xiamen 361005, China
[8]BGT Materials Limited, Photon Science Institute, University of Manchester, Manchester, M13 9PL, UK
[9]Xiamen University Malaysia Campus, 439000 Sepang, Selangor, Malaysia

E-mail: Jin-Cheng Zheng (jczheng@xmu.edu.cn) and K S Novoselov (kostya@manchester.ac.uk)

**Abstract.** Two-dimensional materials are characterised by a number of unique physical properties which can potentially make them useful to a wide diversity of applications. In particular, the large thermal conductivity of graphene and hexagonal boron nitride has already been acknowledged and these materials have been suggested as novel core materials for thermal management in electronics. However, it was not clear if mass produced flakes of hexagonal boron nitride would allow one to achieve an industrially-relevant value of thermal conductivity. Here we demonstrate that laminates of hexagonal boron nitride exhibit thermal conductivity of up to 20 W/m·K, which is significantly larger than that currently used in thermal management. We also show that the thermal conductivity of laminates increases with the increasing volumetric mass density, which creates a way of fine tuning its thermal properties.

## 1. Introduction

Increasing circuit density and miniaturization of modern electronics means that highly efficient heat removal and dissipation is being ever more critical for the reliable operation of electronic devices and systems. Hence, the industry is in urgent need of novel thermally conductive materials suitable for various thermal management applications [1, 2]. It is especially beneficial if such materials are electrically insulating since it would make it possible to apply them directly onto the electronic circuitry. Unfortunately, most of the economically viable insulating materials are characterized by low thermal conductivity, which seriously limits their application as efficient heat spreaders.

It has long been known that bulk hexagonal boron nitride (hBN) possess one of the highest basal plane thermal conductivities among other materials (up to 400 W/m·K at room temperature) [3] and almost matches that of copper and silver [4]. The more recent interest in hBN has been motivated by the search for an electrically insulating counterpart of graphene suitable for thermal management applications [5, 6]. Apart from excellent dielectric properties, few atomic layer hBN crystals demonstrated high values of thermal conductivity approaching that of the bulk value [7-9]. Considering the rare combination of the electrical insulating behaviour with exceptionally high thermal conductivity, hexagonal boron nitride is a promising candidate for the next-generation of thermal management material. However, despite these remarkable properties few-layer hBN crystals are unsuitable for many practical applications which require thermally conductive layers to be either flexible or conformal with the surface, and to have larger thickness to ensure higher thermal diffusivity. All these requirements can be satisfied by obtaining laminates consisting of thin (preferably several monolayers) hBN crystals.

It has been previously demonstrated that graphene laminates possess a relatively high thermal conductivity (up to ~100 W/m·K) along with perfect coating properties [10]. As mentioned previously, the number of potential thermal management applications of such graphene laminates is limited by their high electrical conductivity. On the other hand, hBN laminates are excellent electrical insulators [11]. This property paired with high thermal conductivity can potentially become a paradigm changer for the electronic industry.

In this report we demonstrate that hBN inks can be utilised for the scalable production of both free-standing hBN membranes and surface-supported laminates with high thermal conductivity. The in-plain thermal conductivity of such insulating laminates was found to be as large as 20 W/m·K, which is by an order of magnitude larger than that of the traditionally used heat spreading materials [1, 2].

## 2. Methods

The examined laminates were produced by vacuum filtration of the hBN solution prepared using liquid exfoliation in isopropanol [12]. Upon filtering the substrate with the deposited hBN ink has been dried in a circulation oven at 100°C for 1 hour, and the resulting film has been mechanically delaminated from the substrate for further sample preparation. The thickness of laminate films used in this study varied between 10 μm to 100 μm. The Scanning Electron Microscopy (SEM) image of a typical laminate film is shown in Fig. 1. Analysis of the top and cross-sectional SEM images of the produced films revealed the dominant lateral dimensions of the hBN flakes to be around 1 μm with an average thickness of about 10 nm. In order to obtain various densities of hBN laminate films, some of the samples were subjected to further rolling compression procedure.

The thermal conductivity $\kappa$ of the investigated laminates has been calculated using equation

$$\kappa = \alpha \rho C_p, \tag{1}$$

here $\alpha$ is the in-plane thermal diffusivity, $\rho$ is the material volumetric mass density and $C_p$ is the specific heat. All three parameters were independently determined in experiment.

The thermal diffusivity $\alpha$ as a function of temperature $T$ has been measured by the laser flash method [13] using a commercially available system (Netzsch LFA 457). To measure the in-plane thermal diffusivity a special sample holder has been used, which accommodates free-standing round hBN membrane samples 22 mm in diameter. A 5 mm size spot at the back side of the sample is flash heated by the laser beam through a hole in the sample holder. The heat diffusion as a function of time is registered by the infrared detector along the top circumference of the membrane at 5 mm to 6 mm from the centre of the sample [14]. To avoid undesirable reflections, the sample and sample holder have been spray coated with graphite paint. During the measurements the sample chamber of the laser flash system was continuously purged with nitrogen gas at the rate of 30 ml/min. The sample specific heat $C_p$ was measured by the differential scanning calorimeter (Netzsch DSC 404 F3) using sapphire as a reference sample. The mass density $\rho$ was estimated by weighing a sample of known dimensions with precision electronic balances.

## 3. Results and discussion

To evaluate the effect of structural composition, we measured the thermal conductivity $\kappa$ as a function of temperature $T$ for four hBN laminates with different volumetric mass densities. As seen from Fig. 2 the thermal conductivity is weakly dependent on temperature and increases with the increasing density. The observed values of the thermal conductivity fall in the range between 10 W/m·K to 20 W/m·K, which is certainly an industrially relevant value [1].

To better understand the influence of material density on thermal conductivity, we studied the dependence of $\kappa$ on $\rho$ at room temperature. The density of the laminate samples was controlled in two different ways: (i) by using hBN flakes of different thickness (only limited variations of $\rho$ could be achieved in this way) and (ii) by variation of the additional roller compression applied during preparation of the laminates [10]. Both methods had the same effect on the thermal conductivity. The combined results of this study are presented in Fig. 3. Similarly to the data shown in Fig. 2, the thermal conductivity tends to increase with the increasing density of the laminate. After systematic SEM examination of the laminate samples, we concluded that the mass density variation is mostly due to differences in the flakes packing density, which is directly related to the size of empty voids present between stacked flakes. The schematic representation of two laminates with different $\rho$ is given in Fig. 4. We attribute the reduced thermal conductivity to the discontinuity in the thermal path brought about by larger numbers of empty voids.

To confirm our assumption, we carried out modelling of the thermal flow in laminates with voids. Our numerical simulation was done using ABAQUS 2011 finite element analysis software package. In order to explore the relation between the effective thermal conductivity and the density of hBN laminates we simulated the steady-state heat transfer governed by equation

$$\rho C_p \frac{\partial T}{\partial t} = \frac{\partial}{\partial x}\left(\kappa(T)\frac{\partial T}{\partial x}\right) + \frac{\partial}{\partial y}\left(\kappa(T)\frac{\partial T}{\partial y}\right) + \frac{\partial}{\partial z}\left(\kappa(T)\frac{\partial T}{\partial z}\right) + Q, \tag{2}$$

where *Q* is the heat flux and ∂*T*/∂*t* = 0 (steady-state heat transfer). The modelled system was evaluated with the ABAQUS element type DC2D8 and represented by a strip of orderly stacked solid blocks of thermally conductive media with a lateral size of 1 μm × 1 μm and thickness of 10 nm, Fig. 4. To mimic the hBN flakes, the thermal conductivity of the solid blocks was chosen to be 390 W/m·K at room temperature [7]. The effective density of the modelled laminates was set by changing the overlap area of the adjacent blocks as illustrated in Fig. 4 A and B. Additionally, to account for the imperfect thermal contact between the stacked flakes, a finite thermal contact conductance has been introduced to the model. The resulting effective thermal conductivity $\kappa_{eff}$ of the hBN laminate was calculated using the Fourier law

$$\kappa_{eff} = q \frac{L}{\Delta T}. \tag{3}$$

Here *q* is the total net heat flux through the cross section of the laminate, *L* is the total length of the laminate strip and Δ*T* is the temperature difference between hot and cold ends of the strip. The results of numerical simulation were matched to the experimental data by variation of the thermal contact conductance in the range of $10^5$ W/m²·K to $10^6$ W/m²·K.

The outcome of the simulation is shown by solid curves in Fig. 3. Each of the curves represents the effective thermal conductivity of the laminate with different thermal contact resistances between the stacked hBN flakes. The simulation shows only qualitative agreement with the experimental data because of the simplicity of our model. A more accurate modelling would have to take into account size distribution of the flakes as well as the dependence of the contact conductance on the packing density. Nevertheless, our initial assumption that the thermal conductivity is restricted by the presence of the empty voids inside the laminate has been confirmed by this simple model. Furthermore, it provided a rough estimate of the thermal contact conductance to be of the order $10^6$ W/m²·K. There is no data on the thermal contact conductance available for such systems, however, experimental studies of a rather similar graphene/hBN interface gives a value of around $7 \cdot 10^6$ W/m²·K [15], which is almost an order of magnitude higher than estimated in our simulation. The most probable explanation to this is that the hBN flake surfaces are contaminated with solvent residue, which subsequently reduces thermal conductivity across the flake-to-flake interface.

## 4. Conclusions

In conclusion, we demonstrated that hBN inks can be used to produce laminates with the thermal conductivity as high as 20 W/m·K, which is significantly larger than that for materials currently used in thermal management. We showed that the effective thermal conductivity can be adjusted by varying the laminate packing density. We also identified a potential way to increase the thermal conductivity by improving the quality of the flake-to-flake interface. Being electrically insulating, hBN laminates can potentially overcome many of the problems associated with electrically conductive materials and open a new era in advanced thermal management materials.

## 5. Acknowledgments

This work was supported by the EU FP7 Graphene Flagship Project 604391, ERC Synergy Grant, Hetero2D, EPSRC (Towards Engineering Grand Challenges program), the Royal Society and US Army Research Office. S.V.M. was supported by NUST "MISiS" (grant K1-2015-046) and RFBR (15-02-01221 and 14-02-00792). J.C.Z. and H.Q.W. are supported by National Natural Science Foundation of China (Grant Nos. 11204253 and U1232110).

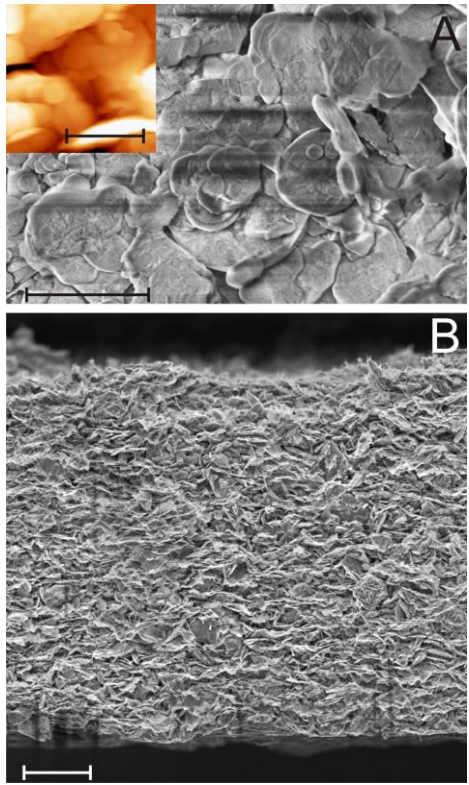

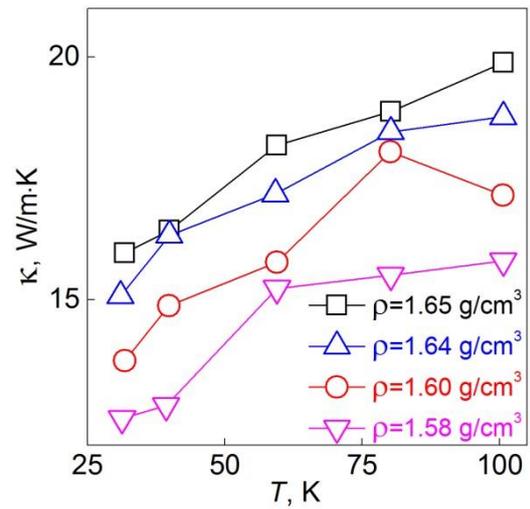

Figure 2. Thermal conductivity $\kappa$ as a function of temperature $T$ measured for different values of hBN laminates density $\rho$.

Figure 1. (A) SEM micrograph of the surface of the hBN laminate. Vertical variations of contrast are due to the charging. Inset: AFM image of the sample (scale black to white is 200nm, giving the average thickness of the flakes ~10nm). Scale bars 1 μm. (B) Cross-sectional SEM image of hBN laminate. Scale bar 10 μm.

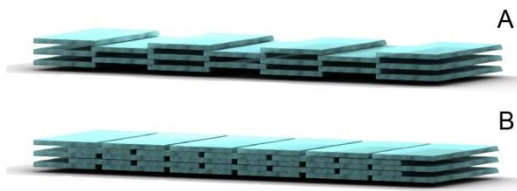

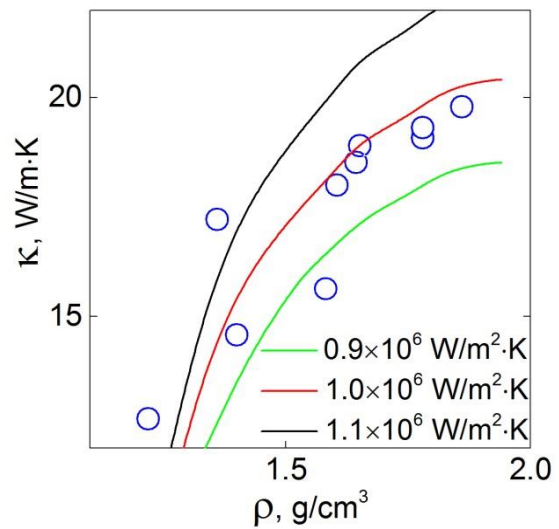

Figure 4. Schematic representation of the laminate model used in numerical simulations for low (A) and high (B) density samples. An individual hBN flake is modelled by a solid block with lateral dimensions 1 μm × 1 μm and thickness 10 nm.

Figure 3. Thermal conductivity $\kappa$ of hBN laminates as a function of density measured at 80°C (blue circles). Solid curves represent results of numerical simulations at different values of the thermal contact conductance (see text for more information).

6